\input{aipcheck}

\documentclass[
    ,final            
  ]
  {aipproc}

\layoutstyle{6x9}

\newcommand{\msun}{\ensuremath{{{\rm M}}_{\scriptscriptstyle \odot}}}
\def\aap{A\&A}
\def\apj{ApJ}
\def\aapr{A\&A Rev.}
\def\apjl{ApJ}
\def\mnras{MNRAS}
\def\araa{ARA\&A}

\def\nat{Nat}

\begin{document}

\title{The first massive black holes}


\author{Marta Volonteri}{
  address={Institut d'Astrophysique de Paris, Paris, France}
}

\begin{abstract}
 I briefly outline recent theoretical developments on the formation of the first massive black holes (MBHs) that may grow into the population of MBHs powering quasars and inhabiting galactic centers today. I also touch upon possible observational tests that may give insights on what the properties of the first MBHs were. 
  
  \end{abstract}

\maketitle


\section{Introduction}
We can trace the presence of MBHs at early cosmic times, as the engines powering the luminous quasars that have been detected at high redshifts,  corresponding to  about a billion years  \citep[$z\simeq 6$;][]{Fanetal2001b,Willott2010} or even 800 million years \citep[$z\simeq 7$;][]{Mortlock2011} after the Big Bang. The evolution of the MBH population, as traced by nuclear activity,  at $z<5-6$ is becoming more strongly constrained by observations in a wide range of wavebands \citep[and references therein]{Merloni2012}.  In the nearby Universe we are not limited to studying active MBHs. In fact the MBHs that we can study best, measuring accurately their masses, are quiescent MBHs, including Sgr A$^*$.

\begin{figure}
  \includegraphics[height=.5\textheight]{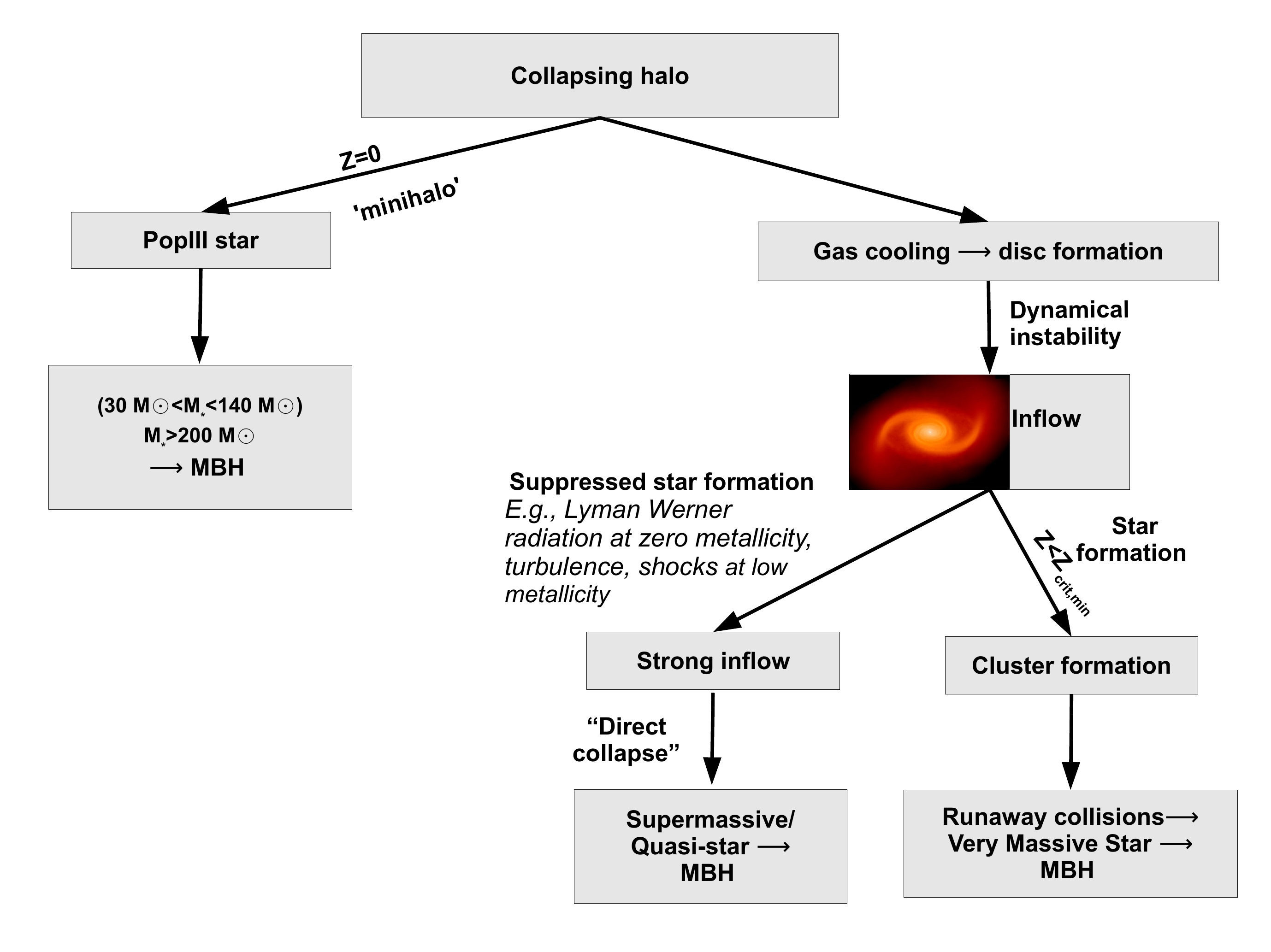}
  \caption{Evolutionary paths that can lead to MBH formation in a proto-galaxy. See  \cite{rees1978} for the seminal version of this type of ``flow-charts".}
\end{figure}

Thus, we now do know that MBHs {\it are} here, but we do not know how they {\it got} here. The early evolution of MBHs, and most notably, what physical mechanism is responsible for their formation are still unknown.   The "flow chart" presented by \cite{rees1978} still stands as a guideline for the possible paths leading to formation of MBH seeds in the center of galactic structures.  In this paper I briefly review the current state-of-the-art in the theoretical modeling of MBH formation (Fig.~1 and~2). See also \cite{Volonteri2010AARV,2012arXiv1203.6075H} for extended reviews.

\section{Formation mechanisms}
As a starting point, let me elaborate on the definition of what a MBH is. Observationally, MBH masses span the range  $\sim 10^5-10^{10}\,\msun$ \citep{Peterson2005,McConnell2011}. These are the masses of MBHs powering AGN and quasars, and also the mass range covered by those MBHs, be them active or not, for which we have direct mass measurement.  Little or no evidence exists of a population of MBHs with lower masses  \cite[but see][]{Farrell09}, until we reach the realm of stellar mass black holes, with masses below a few tens of solar masses. For simplicity, let me here define MBHs as those systems that are too massive to be the end-point of the life of a normal star today, so MBHs are those with mass $>$ 100 \msun. In general, therefore, a stellar black hole can become a MBH if it accretes enough mass. 

\subsection{Remnants of zero-metallicity stars}
One of the first mechanisms proposed to form MBHs at high redshift relates them to the first generation of  stars (Pop III). Low-efficiency cooling in the absence of metals led many authors to suggest that PopIII  stars had high masses, hundreds of solar masses \cite{Couchman86,Abel2000,Bromm04}. In particular, if stars of primordial composition existed with masses greater than $\sim$200 \msun, they are predicted to directly collapse into a black hole of $\sim 100$
\msun \cite{Bond1984,Heger2003}.   
Recent simulations, however, suggest that Pop III stars may have much lower masses \cite{Turk2009,Clark11,Stacy12,Greif11} probably leaving behind remnants with masses similar to today's stellar mass black holes.  It is unlikely that such low-mass black holes can grow to become MBHs to explain the population that we observe powering high redshift quasars or sitting in today's galaxy bulges. A light black hole would dynamically interact with stars of similar mass in the galaxy,  drifting within the galaxy, rather than settling at the center of the galaxy's potential well, in the most favorable location for accreting gas and foster its growth. 
A twist on this scenario may occur if dark matter, in the form of weakly interacting massive particles,  affects the evolution of the first stars \citep{Ripamonti2007,Freese2008,Iocco2008}. If dark matter can reach high densities in the center of the halos where the first stars form, dark matter annihilation may release enough energy to support the star. A star supported by annihilation rather than nuclear fusion may reach  $\sim 500-1000$  $\msun$, and  collapse into MBHs. 

\subsection{Gas-dynamical instability at zero metallicity}
Several works have recently discussed how a rapid inflow of low-angular momentum gas may collapse to form an MBH directly, possibly via a supermassive star phase \cite{LoebRasio1994,Eisenstein1995,OhHaiman2002,BrommLoeb2003,BVR2006,LN2006,Regan2009}. Three conditions must be met. First, enough angular momentum must be shed that rotational support does not halt the collapse of gas. Both global dynamical instabilities \cite{BVR2006} and local dynamical instabilities \citep{LN2006}, have been advocated to trigger such collapse transporting gas inwards and angular momentum outwards. Second, fragmentation and copious star formation must not occur. This to consent enough gas to accumulate and go into MBH formation.  Preventing fragmentation generally requires inhibiting cooling, for instance by requiring no metals and preventing the formation of H$_2$ via a substantial Lyman-Werner background \cite{Dijkstra2008,Shang2010,Petri2012,Agarwal}.   Finally, the process of inflow and MBH formation must occur on timescales shorter than stellar evolution. If  a supernova explodes in a high-redshift protogalaxy it is likely to severely disrupt the whole system, possibly blowing away most, if not all, its gas. 

If these conditions are met, and the gas infall rate is above 1 \msun yr$^{-1}$, the collapsing gas traps its own radiation and forms a quasistatic, radiation pressure-supported supermassive star,  up to $10^6$ \msun \cite{BVR2006,Begelman2007,Begelman2010}. When its core collapses, it forms a black hole. This black hole is confined in the deep core of  the massive radiation-pressure supported envelope created by the inflow, and it grows swallowing it (`quasistar'). The Eddington limit is that for the massive gaseous envelope, thus the black hole can  accrete at highly super-Eddington rate without violating the Eddington luminosity criterion until the black hole mass is about 10\% of the envelope \cite{Ball2011,Begelman2010}

\subsection{Gas-dynamical instability at any metallicity}
Global dynamical instabilities, such as the ``bars within bars''  instability, may act to repeatedly transport gas inwards on the order of a dynamical time \cite{BVR2006}. Large-scale torques from major merger activity may be the trigger of such instabilities \cite{Mayer2010}.  In this case, turbulence may be the inhibitor of fragmentation, and the requirement of metal-free gas may be relaxed (see also \cite{Begelman2009}).  As long as stars form, but the inflow rate is higher than the star formation rate, and gas can accumulate in the galaxy center at rates above 1 \msun yr$^{-1}$, the conditions that lead to the formation of a supermassive star followed by a quasistar \cite{VB2010} are met. 

\subsection{Gas-dynamical instability at low metallicity}
\cite{Inayoshi2012} propose a new scenario for supermassive star ($10^5$ \msun) formation in galaxies with low metallicity, $<10^{-3}$ solar. 
Recent  simulations \cite{Dekel2009, Brooks2009, Bournaud2011} indicate that the early assembly of galaxies in halos of about $10^{11}$ $\msun$ is fueled by cold and dense gas flows penetrating deep to the center from the cosmic web. When  supersonic cold streams collide near the centers of first galaxies,  shocks develop that heat gas.  The post-shock layer initially cools by emitting Lyman alpha photons, and contracts until a temperature of about 8000 K is reached. At high densities, H$_2$ 
rotovibrational levels are collisionally excited. Enhanced H$_2$ collisional dissociation suppresses  further cooling at lower temperature. The shocked layer fragments into massive clouds, which collapse isothermally by Lyman alpha cooling without subsequent fragmentation. These clouds are expected to develop into supermassive stars that eventually collapse into MBHs, retaining up to 90\% of their mass \cite{Shibata2002}. 

\begin{figure}
  \includegraphics[height=.5\textheight]{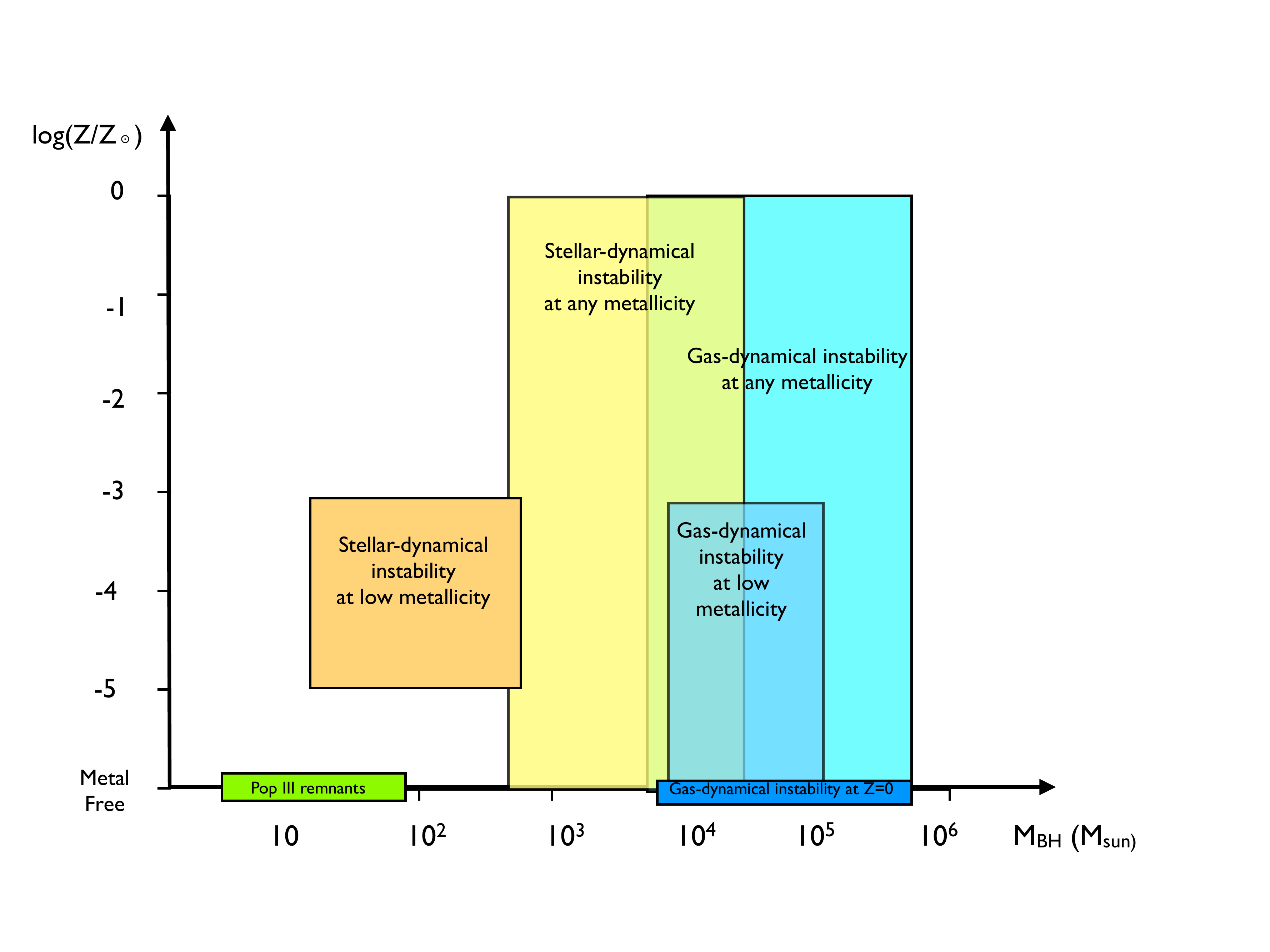}
  \caption{Diagram that shows, qualitatively, the range of metallicities and masses where MBH formation may take place according to the various mechanisms described in this review.}
\end{figure}

\subsection{Stellar-dynamical instability at low metallicity}
\cite{Devecchi2009} propose that MBHs may form in nuclear cluster forming out of the second generation of stars in a galaxy, when the metallicity of the gas is still very low, of order $10^{-5}-10^{-4}$ the metallicity of the Sun (see also \cite{Omukai2008}). In Toomre-unstable proto-galactic discs, instabilities lead to mass infall instead of fragmentation into bound clumps and global star formation in the entire disk. The gas inflow increases the central density, and within a certain compact region star formation ensues and a dense star cluster is formed. At metallicities $\sim 10^{-5}-10^{-4}$ solar, the typical star cluster masses are of order $10^5$ \msun~and the typical half mass radii $\sim 1$ pc \cite{Devecchi2010,Devecchi2012}.  Most star clusters can go into core collapse in $\sim$ 3 Myr (i.e., before any SN explosion takes place), and runaway collisions of stars form a very massive star, leading to an MBH remnant with mass $\sim 10^3$ \msun.  At sub-solar metallicity, the mass loss due to winds is much more reduced in very massive stars, which greatly helps in increasing the mass of the final MBH remnant. 

\subsection{Stellar-dynamical instability at any metallicity}
 \cite{Davies11} recently proposed a scenario for MBH formation triggered by stellar dynamics that does not require low metallicity.  If a galaxy hosts a central nuclear cluster, and strong gas instabilities trigger rapid gas infall, the potential well of the cluster will deepen rapidly.  If the cluster contains a population of compact objects that have segregated in the cluster core, the timescale for core collapse and merger of the compact objects will be shorter than the timescale for dynamical heating via binaries.  The binding energy of the cluster will be large enough that merging black holes are retained, and dynamical ejections are not important.  The result is a runaway collapse of the compact objects that may form an MBH of up to $10^5$ \msun.

\subsection{Primordial black holes}
Primordial black holes may be formed in the early universe before galaxy formation \cite{Zeldovich1967,Hawking1971,Carr2003}.   The predicted masses range  from the Planck Mass (black holes formed at the Planck epoch) to $\simeq$ \msun (black holes formed at the QCD phase transition) up to $10^5$ $\msun$ \cite{Khlopov}. However,  primordial black holes with an initial mass smaller than about $5 \times 10^{14}$~g are  expected to have already  evaporated due to Hawking radiation. For larger masses, that may be of relevance as seeds of MBHs, microlensing  \cite{Alcock2000,Tisserand2007} and spectral distortions of the cosmic microwave background \cite{Ricotti2008}  limit the mass to below $\sim 10^3$ \msun. 

\section{Observational perspectives}
I will not hide the truth that constraining the mechanism of MBH formation is not a trivial task. Neither for theorists trying to propose observational tests, nor for observers going after them. 

\subsection{Direct detection of supermassive stars and quasistars}
Quasistars end their  life with a photospheric temperature of about 3000-4000 K.  Modelling the emission as a blackbody, a quasistar is expected to resemble a featureless red giant with approximately the luminosity of a Seyfert galaxy.  If quasistar formation is  triggered by gas-rich major mergers, however, we expect that the starburst occurring throughout the galaxy may overshine the quasistar, hindering their detection. \cite{VB2010} calculate number counts in the 2--10 $\mu$m band of the {\it James Webb Space Telescope} ({\it JWST}). With optimistic assumptions, the {\it JWST} could detect up to a few quasistars per field. 
\cite{Johnson2009}  suggest that the equivalent width of the 1640 $\AA$ He II line  can be used as an indicator of the initial mass function of Pop III stars, in particular to test whether there were very massive (100 $\msun$) Pop III stars. However, individual Pop III stars, even very massive ones, will likely not be directly detectable by the JWST.

\subsection{Gravitational waves}
Detection of gravitational waves from seeds merging at the redshift of formation \citep{GW3} is probably one of the best ways to discriminate among formation mechanisms.  A large fraction of coalescences of MBHs in the mass range $10^4-10^7$  \msun would be directly observable by a space-based observatory operating at mHz \citep{Pau2012}   frequency and on the basis of the detection rate, and the mass and redshift distribution of the merging MBHs, constraints can be put on the MBH formation process.  A detector sensitive at Hz frequencies, e.g., the proposed Einstein Telescope  \cite{Freise:2009}, would detect merging MBHs with masses of hundreds or a few thousand solar masses, and it would directly  probe formation scenarios that involve MBHs with masses in this range. 

\subsection{Dwarf galaxies}
We can use simple arguments to predict that the best probe of MBH formation in the local Universe can be found in dwarf galaxies. A MBH hosted by a massive galaxy has a low probability of being ``pristine", as it is likely that it has increased its mass by accretion, or it has experienced mergers and dynamical interactions. 
 Dwarf galaxies undergo instead a quieter  evolution, and as a result, at low masses the MBH occupation fraction and the distribution of MBH masses still retain some ``memory'' of the original seed mass distribution. The signature of the efficiency of the formation of MBH seeds will consequently be stronger in  dwarf galaxies.  The occupation fraction and distribution of MBH masses in dwarf galaxies are possible diagnostics \citep{VLN2008,svanwas2010}.  

\subsection{Wandering MBHs}
If MBHs form at early times, over the complex hierarchical assembly of galaxies a population of wandering black holes may be created in galactic halos \cite{Schneider2002,Islam2003,VolonteriPerna2005,Mapelli2006,Mapelli2008,Bellovary2010}.  Detection of the population of wandering black holes is expected to be challenging, due to the small number of objects, very low predicted luminosity and short activity timescales at high luminosity. Currently, the best candidate of a wandering MBH (not necessarily linked to MBH `seed' formation, though, cf. \cite{PZ2004}) is HLX-1  \cite{Farrell09}.  

\begin{theacknowledgments}
  The author acknowledges funding support from NASA, through award ATP NNX10AC84G; from SAO, through award TM1-12007X, from NSF, through award AST 1107675, and from a Marie Curie Career Integration grant. 
\end{theacknowledgments}

\bibliographystyle{aipproc}   



\end{document}